\documentclass[12pt]{article}
\usepackage[left=2.5cm, bottom=2.5cm, right=2.5cm, top=2.5cm]{geometry}

\usepackage{sectsty}
\usepackage{rotating}
\usepackage{comment}
\usepackage{tikz}
\usepackage{amssymb}
\usepackage{graphicx}
\usepackage{mathbbol}
\usepackage{nicefrac}
\usetikzlibrary{patterns}
\usepackage{subcaption}
\makeatletter
\def\tikz@arc@opt[#1]{
  {%
    \tikzset{every arc/.try,#1}%
    \pgfkeysgetvalue{/tikz/start angle}\tikz@s
    \pgfkeysgetvalue{/tikz/end angle}\tikz@e
    \pgfkeysgetvalue{/tikz/delta angle}\tikz@d
    \ifx\tikz@s\pgfutil@empty%
      \pgfmathsetmacro\tikz@s{\tikz@e-\tikz@d}
    \else
      \ifx\tikz@e\pgfutil@empty%
        \pgfmathsetmacro\tikz@e{\tikz@s+\tikz@d}
      \fi%
    \fi
    \tikz@arc@moveto
    \xdef\pgf@marshal{\noexpand%
    \tikz@do@arc{\tikz@s}{\tikz@e}
      {\pgfkeysvalueof{/tikz/x radius}}
      {\pgfkeysvalueof{/tikz/y radius}}}%
  }%
  \pgf@marshal%
  \tikz@arcfinal%
}
\let\tikz@arc@moveto\relax
\def\tikz@arc@movetolineto#1{%
  \def\tikz@arc@moveto{\tikz@@@parse@polar{\tikz@arc@@movetolineto#1}(\tikz@s:\pgfkeysvalueof{/tikz/x radius} and \pgfkeysvalueof{/tikz/y radius})}}
\def\tikz@arc@@movetolineto#1#2{#1{\pgfpointadd{#2}{\tikz@last@position@saved}}}
\tikzset{%
  move to start/.code=\tikz@arc@movetolineto\pgfpathmoveto,%
  line to start/.code=\tikz@arc@movetolineto\pgfpathlineto}
\makeatother
\makeatletter
\newcommand{\labeltext}[3][]{%
    \@bsphack%
    \csname phantomsection\endcsname
    \def\tst{#1}%
    \def\labelmarkup{}
    \def\refmarkup{}%
    \ifx\tst\empty\def\@currentlabel{\refmarkup{#2}}{\label{#3}}%
    \else\def\@currentlabel{\refmarkup{#1}}{\label{#3}}\fi%
    \@esphack%
    \labelmarkup{#2}
}
\makeatother
\usepackage{pgfplots}
\usetikzlibrary{intersections, pgfplots.fillbetween}
\usetikzlibrary{decorations.pathreplacing}
\usepackage{amsmath}
\usepackage{amsthm}
\usepackage{bm}
\usepackage{multirow}
\usepackage[normalem]{ulem}
\usepackage{pdflscape}
\usepackage{pgfplots}
\usepackage{mathtools}
\usepackage{theoremref}
\usepackage{xcolor}
\usepackage{changepage}
\colorlet{linkequation}{blue}
\usepackage[authoryear,longnamesfirst,round]{natbib}
\setcitestyle{authoryear,open={(},close={)},aysep={},citesep={;}}
\colorlet{linkequation}{blue}
\usepackage[colorlinks=true,allcolors=blue]{hyperref}
\newcommand*{\SavedEqref}{}
\let\SavedEqref\eqref
\renewcommand*{\eqref}[1]{%
  \begingroup
    \hypersetup{
      linkcolor=linkequation,
      linkbordercolor=linkequation,
      breaklinks=true,   
    }%
    \SavedEqref{#1}%
  \endgroup
}
\newcommand\bref[1]{{\hypersetup{linkcolor=blue}\autoref{#1}}}
\makeatletter
\renewcommand\@makefnmark{\hbox{\@textsuperscript{\normalfont\color{black}\@thefnmark}}}
\makeatother
\useunder{\uline}{\ul}{}
\definecolor{gray1}{gray}{0.8}
\definecolor{gray2}{gray}{0.6}
\definecolor{gray3}{gray}{0.4}
\DeclareMathOperator*{\argmax}{\arg\!\max}
\DeclareMathOperator*{\argmin}{\arg\!\min}

\newcommand*\diff{\mathop{}\!\mathrm{d}}

\pgfplotsset{compat=1.17}
\theoremstyle{plain}
\newtheorem{thm}{Theorem}
\theoremstyle{plain}
\newtheorem{prop}{Proposition}
\theoremstyle{plain}

\theoremstyle{definition}
\newtheorem{defi}{Definition}
\theoremstyle{definition}

\theoremstyle{plain}
\newtheorem{lem}{Lemma}
\theoremstyle{definition}
\theoremstyle{plain}

\theoremstyle{remark}

\theoremstyle{remark}
\newtheorem{exa}{Example}
\usetikzlibrary{arrows,shapes,positioning}
\tikzstyle{every lower node part}=[font=5pt]
\sectionfont{\fontsize{14}{16}\selectfont}
\usepackage{setspace}

\usepackage{caption}
\usepackage{subcaption}

\renewcommand{\omega}{x}

\definecolor{BlueGreenT}{rgb}{0.051,0.59,0.729}

\definecolor{GoldOrangeT}{rgb}{0.9215,0.6235,0.1843}

\begin{document}
\title{\textbf{Regulating Oligopolistic Competition}\thanks{This paper was previously circulated with the title ``Efficient Market Structures Under Incomplete Information.'' We are grateful to Florian Ederer, Gary Gorton, Yunzhi Hu, Doron Ravid, Jidong Zhou, and participants at the 2022 AEA Meeting for very helpful feedback and comments.  }
}
\author{
    Kai Hao Yang\thanks{Yale School of Management, Email: kaihao.yang@yale.edu} 
    \and 
    Alexander K. Zentefis\thanks{Yale School of Management, Email: alexander.zentefis@yale.edu} }

\maketitle
\begin{abstract}

We consider the problem of how to regulate an oligopoly when firms have private information about their costs. In the environment, consumers make discrete choices over goods, and minimal structure is placed on the manner in which firms compete. In the optimal regulatory policy, firms compete on price margins, and based on firms' prices, the regulator charges them taxes or give them subsidies, and imposes on each firm a ``yardstick'' price cap that depends on the posted prices of competing firms.

%

\end{abstract}
\begin{flushleft}
\textbf{\vfill{}
JEL classification:} D40, D82, L5 \textbf{}\linebreak{}
\textbf{Keywords:} regulation, price caps, mechanism design\\
\par\end{flushleft}

\vfill{}

\pagebreak{}

\setcounter{footnote}{0}

\begin{spacing}{1.5}

\section{Introduction}

In a seminal paper, \citet{BM82} derive the optimal regulation of a monopolist whose costs are unknown to the regulator. In their model, the monopolist faces a commonly known inverse demand function, and the regulator's problem is to determine whether the monopolist is allowed to produce at all, and if so, how the monopolist's price and transfer should be determined as functions of the production cost the monopolist reports. Baron and Myerson show that the optimal regulatory policy characterizes a price schedule for all monopolist types permitted to operate, where production is permitted only for types for which consumer surplus under the optimal policy exceeds the fixed cost of production.\footnote{See \citet{amador2016regulating} and \cite{guo2019robust} for other recent papers related to \cite{BM82}.}

In this paper, we generalize \citet{BM82} to an oligopoly setting. Firms have independent, one-dimensional private information about their costs, and consumers choose one among many differentiated goods to consume. These two changes to the Baron-Myerson model substantially enriches the regulatory problem. Instead of a one-dimensional demand quantity as in \citet{BM82}, the relevant allocation is the entire distribution of matches between consumers and firms. And with multiple firms, the regulator must now take into account the strategic manner in which firms compete (i.e., the model of market conduct).

To derive the optimal regulatory policy, we follow the Baron-Myerson approach of searching for an incentive-efficient allocation that maximizes a linear social welfare function of consumer surplus and firms' profits. To capture a wide range of manners in which firms compete, we consider all possible indirect mechanisms, each of which consists of arbitrary sets of strategies firms can adopt and an arbitrary mapping from firms' strategy profiles to market outcomes. This expansive treatment accommodates the fact that styles of competition can differ significantly between markets. We demonstrate that the class of indirect mechanisms we search over nests competing on price á la \citet{bertrand1883book}, competing on quantity á la \citet{cournot1838recherches}, competing over differentiated goods \citep{perloff1985equilibrium}, and consumer search \citep{varian1980model,narasimhan1988competitive,armstrong2019discriminating}. Looking across this large class, we prove that every constrained efficient indirect mechanism is equivalent to price competition, but with lump-sum transfers and a certain form of price controls---namely, firm-specific price caps that depend on the prices of competitors (i.e., ``yardstick'' price caps).\footnote{\citet{wang2000regulating} also studies the problem of regulating an oligopoly with unknown costs. The focus there is about quantity regulation, and it assumes homogeneous goods, two firm-cost types, no fixed costs, and a one-dimensional demand curve.}

The intuition behind our main characterization is as follows: Because we presume independent private types across firms, we can adopt the Myersonian approach to keep track of revenues as functions of an allocation of goods to consumers. Doing so implies that the efficient allocation must allocate goods to consumers who have the highest ex-post \emph{virtual surplus}  on the intensive margin (i.e., the difference between a consumer's value for a good and the virtual marginal cost of the firm supplying the good). At the same time, the efficient allocation must also select the set of firms that generate the highest ex-ante virtual surplus on the extensive margin when taking fixed costs into account. Price competition implements the efficient allocation on the intensive margin. By properly designing lump-sum transfers between consumers and firms as functions of firms' prices, one can incentivize firms to post prices that exactly reflect their virtual marginal cost. With virtual marginal costs (and, hence, firms' private types) being reflected in prices, one can then select the most efficient firms based solely on the information collected from posted prices on the extensive margin. This selection criterion leads to the yardstick price caps. 


The rest of this paper is organized as follows: In \bref{sec:model}, we introduce the model, define our indirect mechanisms, and specify the welfare criterion for efficiency. \bref{sec:pryce_cap_efficiency} states the main result, and \bref{sec:proof_thm1} has the proof. \bref{sec:example_and_properties} provides an example of an efficient regulatory policy. \bref{sec:extensions} gives extensions of the baseline model: one incorporating capacity constraints; and the other, consumers' ex-post individual rationality constraints.  \bref{sec:conclusion} concludes. Omitted proofs can be found in the Appendix. 


\section{Model}
\label{sec:model}
\subsection{Primitives}
A number $N \geq 1$ of firms produce $N$ heterogeneous goods. Each firm $i$ has cost function $C_i(q)=\theta_i(q+\kappa_i)$, where $q$ is quantity and $\kappa_i\geq0$ is commonly known. Meanwhile, $\theta_i \geq 0$ represents the firm's cost efficiency and is private information. A lower $\theta_i$ implies that a firm is more cost-efficient. We assume that $\theta=(\theta_i)_{i=1}^N \in \mathbb{R}^N$ is independent and $\theta_i$ follows a distribution $G_i$, which has a support $\Theta_i:=[\underline{\theta}_i,\overline{\theta_i}]$, with $0\leq \underline{\theta}_i \leq \overline{\theta}_i <\infty$.

There is a unit mass of consumers. Each consumer has unit demand and heterogeneous values $\mathbf{v} \in V \subseteq \mathbb{R}_+^N$, so that a consumer with value vector $\mathbf{v}=(v_1,\ldots,v_N)$ has value $v_i$ for firm $i$'s good. The consumers' values are distributed according to a measure $F \in \Delta(V)$. 
\subsection{Indirect Mechanisms}
An indirect mechanism $\mathcal{M}$ is a tuple $\mathcal{M}=(S_i,r_i,\bm{\mu}_i,t_i)_{i=1}^N$ that assigns firms' strategies to (i) market entry probabilities, (ii) an allocation of goods to consumers, and (iii) firm revenues; where, for all $i$, $S_i$ is an arbitrary (measurable) set, $r_i$ is a mapping from $S:=\prod_{i=1}^N S_i$ to $[0,1]$, $t_i$ is a mapping from $S$ to $\mathbb{R}$, and $\bm{\mu}_i$ is a mapping from $V \times S$ to $[0,1]$, such that $\sum_{i=1}^N \bm{\mu}_i(\mathbf{v}|s)\leq 1$ for all $(\mathbf{v},s) \in V \times S$. 

For any indirect mechanism, $S_i$ describes firm $i$'s available strategies (e.g., chosen price, chosen quantity, or entry decision). Given any strategy profile $s \in S=\prod_{i=1}^N S_i$, $r_i(s) \in [0,1]$ denotes the probability that firm $i$ enters the market; $\bm{\mu}_i(\mathbf{v}|s) \in [0,1]$ represents the share of consumers with value $\mathbf{v}$ who receives firm $i$'s good, conditional on firm $i$ being in the market; and $t_i(s)$ is the revenue of firm $i$. We normalize the firms' outside options to zero, and we require that any indirect mechanism must allow an opt-out option $s_0 \in S_i$ such that $t_i(s_0,s_{-i})=\bm{\mu}_i(\mathbf{v}|s_0,s_{-i})=r_i(s_0,s_{-i})=0$ for all $i$, for all $s_{-i} \in S_{-i}$, and for all $\mathbf{v} \in V$. 

Given $\mathcal{M}$, the timing of events is as follows: (1) types $\{\theta_i\}_{i=1}^N$ are drawn independently from $\{G_i\}_{i=1}^N$, and each firm privately observes its own type; (2) firms simultaneously choose $s_i$ from $S_i$; and (3) each firm $i$ receives ex-post payoff 
\[
\pi_i(s,\theta_i|\mathcal{M}):=t_i(s)-r_i(s)\theta_i\left(\int_V\bm{\mu}_i(\mathbf{v}|s)F(\diff \mathbf{v})+\kappa_i\right).
\]
Notice that $\mathcal{M}$ defines a Bayesian game where each firm $i$ has private type $\theta_i \in \Theta_i$, strategy space $S_i$, and payoff function $\pi_i(s,\theta_i|\mathcal{M})$.  

\subsection{Example Indirect Mechanisms}
\label{sec:range_of_games}

Here we provide three example indirect mechanisms: price competition, quantity competition, and consumer search. Other examples are in Online Appendix \ref{A1}.



\begin{exa}[\textbf{Price Competition}]\label{ex2}
The following mechanism describes a price competition model. Under this mechanism, all $N$ firms operate in the market and compete on the price margin (i.e., each firm $i$ sets price $s_i \geq 0$). After seeing firms' prices $s=(s_1,\ldots,s_N)$, a consumer buys from the firm providing the highest surplus. 

Specifically, each firm $i$ has strategy space $S_i=\mathbb{R}_+$. Under any strategy profile $s \in S$, firm $i$'s entry probability is $r_i(s)=1$ and revenue is $t_i(s)=s_i\int_V\bm{\mu}_i(\mathbf{v}|s)F(\diff \mathbf{v})$, where $\bm{\mu}_i$ is given by
\[
\bm{\mu}_i(v|s)=\left\{\begin{array}{cc}
\frac{1}{|\mathbb{M}(\mathbf{v},s)|},&\mbox{if } v_i-s_i=\max_{j}\{v_j-s_j\}\mbox{ and } v_i \geq s_i\\
0,&\mbox{otherwise}
\end{array}\right.,
\] 
for all $i \in \{1,\ldots,N\}$ and for all $s \in S$, with $\mathbb{M}(\mathbf{v},s):=\argmax_i\{v_i-s_i\}$.\footnote{Notice that with different specifications of the value distribution $F$, this mechanism corresponds to various canonical competition models. In particular, by assuming that $\mathbf{v}$ is perfectly correlated (i.e., $v_1=\ldots,v_N=v$ with $F$-probability 1), we have the classical Bertrand competition model \citep{bertrand1883book}, but with private marginal costs; by assuming that $\mathbf{v}$ is independent, we have the model \'{a} la \citet{perloff1985equilibrium}; by assuming that $N=2$ and that $\mathbf{v}$ is perfectly negatively correlated (i.e., $v_1+v_2=1$ with $F$-probability 1), we have the Hotelling location model \citep{hotelling1929}.}
\end{exa}
 
\begin{exa}[\textbf{Quantity Competition}]
Suppose that $F$ is atomless. Then there exists an indirect mechanism that describes quantity competition of which the classical Cournot model \citep{cournot1838recherches} is a special case.  Under this mechanism, each firm $i$ chooses quantity $s_i \in [0,1]$ it wishes to sell. Market prices (and, hence, the allocation of goods) are determined through a system of inverse demand functions $\{\bm{p}_i\}_{i=1}^N$. For any $i$, $\bm{\mu}_i$ is defined so that firm $i$ sells $s_i$ units at price $\bm{p}_i(s)$ if $\sum_j s_j \leq 1$, and sells $\frac{s_i}{\sum_j s_j}$ units at price 0 if $\sum_j s_j >1$. This mechanism is strategically equivalent to a quantity competition game with inverse demand functions $\{\bm{p}_i\}_{i=1}^N$.\footnote{See more details in Online Appendix \ref{A1}.}
\end{exa} 

\begin{exa}[\textbf{Consumer Search and Promotional Sales}] 
Consider the price competition mechanism given by \bref{ex2}, except that $\bm{\mu}_i$ becomes 
\[
\bm{\mu}_i(\mathbf{v}|s)=\left\{
\begin{array}{cc}
\gamma_i+\left(1-\sum_{j=1}^N \gamma_j\right)\frac{\mathbf{1}\{s_i \in \mathbb{M}(\mathbf{v},s)\}}{|\mathbb{M}(\mathbf{v},s)|},& \mbox{if } v_i \geq s_i\\
0,& \mbox{if } v_i<s_i
\end{array}
\right..
\]
This indirect mechanism then describes a model with ``captive consumers'' and ``shoppers,'' where each firm $i$ has $\gamma_i \in [0,1]$ share of captive consumers who can only see its price, while the remaining consumers can visit all firms and see all firms' prices.\footnote{Notice that if $\mathbf{v}$ is perfectly correlated so that with $F$-probability 1, $v_1=v_2=\cdots=v_N$, this mechanism describes the promotional sales model of \citet{armstrong2019discriminating}, which in turn nests the consumer search model of \citet{varian1980model} and  \citet{narasimhan1988competitive}.}
\end{exa}

\paragraph{Discussion.}As hinted by the examples above, \emph{any} Bayesian game that models competition among $K \leq N$ firms is included in the class of indirect mechanisms we consider. As such, our analysis of mechanisms applies to all possible static models of competition with fixed preferences and technology, regardless of a model's assumptions about firm conduct, market power, or price determination. Any dynamic model that can be represented in strategic form is also eligible, as are markets in which prices are determined via bilateral bargaining.

Of course, not all competitive games among $K\leq N$ firms have an equilibrium. Furthermore, even if an equilibrium exists, some equilibria might be extremely difficult to characterize. A benefit of our framework is that, as explained below, it bypasses explicit characterizations of equilibria and only focuses on the outcomes. Across this broad range, our main interest is to characterize the efficient indirect mechanisms and explore ways to implement them. To this end, we first formally define our notion of efficiency.      

\subsection{Defining Efficiency}

For any indirect mechanism $\mathcal{M}=(S_i,r_i,\bm{\mu}_i,t_i)_{i=1}^N$, and for any Bayes-Nash equilibrium $\sigma=\prod_{i=1}^N \sigma_i$ of the induced Bayesian game, where $\sigma_i:\Theta_i \to \Delta(S_i)$ is firm $i$'s equilibrium strategy, let 
\[
\Pi_i(\theta_i|\mathcal{M},\sigma):=\mathbb{E}_{\theta_{-i}}\left[ \int_S\pi_i(s,\theta_i|\mathcal{M})\sigma(\diff s|\theta)\right]
\]
denote firm $i$'s interim profit, and let 
\[
\Sigma(\mathcal{M},\sigma):=\mathbb{E}_\theta\left[\int_{V\times S} \sum_{i=1}^N v_i\bm{\mu}_i(\mathbf{v}|s)\sigma(\diff s|\theta)F(\diff \mathbf{v})-\sum_{i=1}^N \int_S t_i(s)\sigma(\diff s|\theta)\right].
\]
denote the expected consumer surplus. With this notation, we have the following definition of Pareto dominance:

\begin{defi}
An indirect mechanism $\mathcal{M}$ and a Bayes-Nash equilibrium $\sigma$ \emph{dominates} another indirect mechanism $\mathcal{M}'$ and Bayes-Nash equilibrium $\sigma'$ if 
\[
\Sigma(\mathcal{M},\sigma) \geq \Sigma(\mathcal{M}',\sigma')
\]
and
\[
\Pi_i(\theta_i|\mathcal{M},\sigma) \geq \Pi_i(\theta_i|\mathcal{M}',\sigma')
\]
for all $i$ and for all $\theta_i \in \Theta_i$, with at least one inequality being strict. 
\end{defi}

We can then define (constrained) efficiency in the usual sense:

\begin{defi}
An indirect mechanism $\mathcal{M}$ is (constrained) \emph{efficient} if there exists a Bayes-Nash equilibrium $\sigma$ in the Bayesian game induced by $\mathcal{M}$ such that no other indirect mechanisms and Bayes-Nash equilibria dominate $\mathcal{M}$ and $\sigma$.
\end{defi}

Just as in \citet{BM82}, consumers are not explicitly included as agents in an indirect mechanism. Rather, they are implicitly embedded into the allocation rules $\bm{\mu}$ (just as they are embedded into the market demand in \citealp{BM82}). A consequence of this formulation is that consumers do not have participation constraints. Thus, there might be indirect mechanisms that leave consumers with negative surplus while granting firms unbounded revenue via taxation and subsidy. To rule out these trivial cases, we focus hereafter on mechanisms that leave consumers with non-negative surplus. One way to incorporate this constraint, just as in \citet{BM82}, is to represent efficiency with a social planner maximizing a weighted sum of consumer surplus and firm profits, and, on average, assigning a relatively higher weight to consumers than to firms. We summarize this observation in the following lemma:

\begin{lem}\label{pareto}
An indirect mechanism $\mathcal{M}$ is (constrained) efficient with consumers obtaining non-negative surplus if and only if there exists a collection of nondecreasing, right-continuous functions $\{\Lambda_i\}_{i=1}^N$ on $\Theta_i$, with $0\leq \Lambda_i(\theta_i) \leq G_i(\theta_i)$ such that for any indirect mechanism $\mathcal{M}'$ and Bayes-Nash equilibrium $\sigma$',
\begin{equation}\label{planner}
\Sigma(\mathcal{M},\sigma)+\sum_{i=1}^N \int_{\Theta_i} \Pi(\theta_i|\mathcal{M},\sigma)\Lambda_i(\diff \theta_i)\\
\geq \Sigma(\mathcal{M}',\sigma')+\sum_{i=1}^N \int_{\Theta_i} \Pi(\theta_i|\mathcal{M}';\sigma')\Lambda_i(\diff \theta_i).
\end{equation}
\end{lem}

In essence, \bref{pareto} uses the familiar method that represents the Pareto frontier with solutions of a planner's problem, where the planner maximizes a weighed sum of consumer surplus and firm profits. The Pareto weights for consumers are normalized to 1, whereas the weights for firm $i$ are given by $\{\Lambda_i(\theta_i)\}_{\theta_i \in \Theta_i}$.\footnote{This is because the dominance criterion is applied in the interim stage for each realization of types $\theta_i$.} The restriction $\Lambda_i(\theta_i) \leq G_i(\theta_i)$ ensures that firms receive lower weights then consumers do. In the special case where $N=1$ and $\Lambda_i(\theta_i)=(1-\alpha)G_i(\theta_i)$, \eqref{planner} exactly matches the regulator's objective in \citet{BM82}. In other words, the restriction that $\Lambda_i(\theta)_i \leq G_i(\theta_i)$ can be regarded as a multi-firm and interim analog of the parameterization with $\alpha \in [0,1]$ used in \citet{BM82}.

\section{Efficiency of PRYCE CAP Mechanisms}
\label{sec:pryce_cap_efficiency}
In what follows, we present our main result. As noted in the previous section, our central interest is in characterizing the efficient mechanisms and exploring practical ways to implement them. Although there are infinitely many possible mechanisms and some of them can be extremely complex, we show that the efficient ones are ``simple,'' in the sense that any efficient indirect mechanism is equivalent to one that belongs to a natural class. This class of mechanisms involves price competition with lump-sum transfers and firm-specific price caps that depend on the chosen prices of competitors. We call these price caps \emph{yardstick price caps}, and we refer to this class as PRYCE CAP mechanisms, which we define next.\footnote{In naming the price caps, we use the word ``yardstick'' in a way similar to \citet{shleifer1985theory}'s use of the word, in that a firm's regulation depends on characteristics of other firms.}  

\begin{defi}
\label{def:pryce_cap}
$\mathcal{M}=(S_i,r_i,\bm{\mu}_i,t_i)_{i=1}^N$ is a price competition mechanism with lump-sum transfers and yardstick price caps (PRYCE CAP)  if, for any $i$,
\begin{enumerate}
\item $S_i=\mathbb{R}_+$. 
\item For any $s \in S$, $r_i(s)=\mathbf{1}\{s_i \leq \bar{p}_i(s_{-i})\}$,
for some $\bar{p}_i:S_{-i} \to \mathbb{R}_+\cup \{\infty\}$.
\item For any $\mathbf{v} \in V$, and for any $s \in S$, 
\[
\bm{\mu}_i(\mathbf{v}|s)=\left\{
\begin{array}{cc}
1,& \mbox{if } v_i-s_i>\max_{\{j|r_j(s)=1,\, j \neq i\}} (v_j-s_j)^+ \mbox{ and } r_i(s)=1\\
0,& \mbox{if } v_i-s_i<\max_{\{j|r_j(s)=1, \, j \neq i\}} (v_j-s_j)^+ \mbox{ or } r_i(s)=0
\end{array}
\right..
\]
\item For any $s \in S$, $t_i(s)=s_i\int_V\bm{\mu}_i(\mathbf{v}|s)F(\diff \mathbf{v})-\tau_i(s_i)$, for some $\tau_i:S_i \to \mathbb{R}$. 
\end{enumerate}
\end{defi}

Under a PRYCE CAP mechanism, each firm $i$ simultaneously announces a price $s_i \geq 0$. Given the announced prices $s=(s_1,\ldots,s_N)$, a firm is first selected into the market based on whether its announced price $s_i$ is below its price cap $\bar{p}_i(s_{-i})$. The price caps and the rules for market entry are thus intimately linked. When choosing a price to announce, a firm accounts for both its own price cap and the effect that its choice will have on the price caps of other firms. Among the firms that enter the market, consumers then see the announced prices and decide whether to purchase and which firm to buy from. Finally, each firm is compensated or taxed via lump-sum transfers from consumers. This transfer amount $\tau_i(s_i)$ depends only on a firm's own price.

Notice that if $\bar{p}_i(s_{-i})=\infty$ and $\tau_i(s_i)=0$ for all $i$ and for all $s$, a PRYCE CAP mechanism reduces to a pure price competition model (see \bref{ex2} above). From this perspective, PRYCE CAP mechanisms can be regarded as generalizations of pure price competition models that are commonly assumed, with the differences being lump-sum transfers $\{\tau_i\}_{i=1}^N$ and yardstick price caps $\{\bar{p}_i\}_{i=1}^N$. 


With the formal definition of PRYCE CAP mechanisms presented, we now state our main result. 
 
\begin{thm}\label{thm1}
Any efficient indirect mechanism is equivalent to a PRYCE CAP mechanism. 
\end{thm}

The significance of \bref{thm1} is that, among infinitely many indirect mechanisms, the efficient ones are equivalent to a PRYCE CAP mechanism. This means that price competition, together with interventions solely in the form of lump-sum transfers and price ceilings, are enough to achieve constrained Pareto efficiency. PRYCE CAP mechanisms emerge as efficient out of an expansive set of market environments in which firms and consumers partake, with each environment potentially experiencing enormously complicated forms of firm conduct, barriers to entry, and regulatory policies.

Furthermore, \bref{thm1} implies that omniscient knowledge about the market setting is not required to implement an efficient regulation. More precisely, implementing a PRYCE CAP mechanism does not require knowledge about each individual consumer's value vector $\mathbf{v}$. With proper lump-sum transfers and yardstick price caps, firms would post correct prices and consumers would sort themselves into the efficient allocation.


The fact that any efficient regulation is equivalent to a PRYCE CAP mechanism sheds light on which regulatory policies are necessary and which are not. After all, \bref{thm1} implies that if firms compete on price, any regulatory policy other than lump-sum transfers and yardstick price caps is unwarranted for reaching efficiency in an environment like ours. In other words, lump-sum transfers and price ceilings can be regarded as \emph{minimal} regulatory policies, given that firms are able to compete (only) on the price margin.



\section{Proof of Theorem 1}
\label{sec:proof_thm1}

This section provides the proof of \bref{thm1}. First, notice that by the revelation principle (\citealp{M79}), it is without loss to restrict attention to incentive compatible and individually rational direct mechanisms. A direct mechanism is a mechanism $(S_i,r_i,\bm{\mu}_i,t_i)_{i=1}^N$ where $S_i=\Theta_i$ for all $i$. For simplicity, we refer to a direct mechanism as a mechanism, and we denote it by $(r,\bm{\mu},t)$ hereafter when there is no confusion. 

Specifically, a mechanism is incentive compatible if, for all $i$ and for all $\theta_i,\theta_i' \in \Theta_i$, 
\begin{align*}\label{ic}
\mathbb{E}_{\theta_{-i}}&\left[t_i(\theta_i,\theta_{-i})-r_i(\theta_i,\theta_{-i})\theta_i\left(\int_V\bm{\mu}_i(\mathbf{v}|\theta_i,\theta_{-i})F(\diff \mathbf{v})+\kappa_i\right)\right] \\
&\geq \mathbb{E}_{\theta_{-i}}\left[t_i(\theta_i',\theta_{-i})-r_i(\theta_i',\theta_{-i})\theta_i\left(\int_V\bm{\mu}_i(\mathbf{v}|\theta_i',\theta_{-i})F(\diff \mathbf{v})+\kappa_i\right)\right],\tag{IC}
\end{align*}
and is individually rational if, for all $\theta_i \in \Theta_i$,
\begin{align*}\label{ir}
\mathbb{E}_{\theta_{-i}}\left[t_i(\theta_i,\theta_{-i})-r_i(\theta_i,\theta_{-i})\theta_i\left(\int_V\bm{\mu}_i(\mathbf{v}|\theta_i,\theta_{-i})F(\diff \mathbf{v})+\kappa_i\right)\right] \geq 0. \tag{IR}
\end{align*}

Under any incentive compatible and individually rational mechanism $(r,\bm{\mu},t)$, firm $i$'s interim expected profit is 
\[
\Pi_i(\theta_i|r,\bm{\mu},t)=\mathbb{E}_{\theta_{-i}}\left[t_i(\theta_i)-r_i(\theta)\theta_i\left(\int_V\bm{\mu}_i(\mathbf{v}|\theta)F(\diff \mathbf{v})+\kappa_i\right)\right],
\]
while the expected consumer surplus is 
\[
\Sigma(r,\bm{\mu},t):=\mathbb{E}_{\theta}\left[\sum_{i=1}^Nr_i(\theta)\int_V\bm{\mu}_i(\mathbf{v}|c)v_iF(\diff \mathbf{v})-\sum_{i=1}^N t_i(\theta)\right].
\]

As a result, by \bref{pareto}, an incentive compatible and individually rational mechanism is efficient if and only if it is the solution to the following problem:
\begin{equation}\label{objj}
\sup_{(r,\bm{\mu},t)} \bigg[\sum_{i=1}^N \int_{\Theta_i} \Pi_i(\theta_i|r,\bm{\mu},t)\Lambda_i(\diff \theta_i)+\Sigma(r,\bm{\mu},t)\bigg], 
\end{equation}
subject to \eqref{ic} and \eqref{ir}, for some collection of nondecreasing and right-continuous functions $\{\Lambda_i\}$ with $0\leq \Lambda_i(\theta_i) \leq G_i(\theta_i)$ for all $\theta_i \in \Theta_i$.

Meanwhile, using the standard envelope arguments, we can characterize incentive compatibility by a revenue equivalence formula and a monotonicity condition, as summarized by the following lemma. 

\begin{lem}\label{rev}
A mechanism $(r,\bm{\mu},t)$ is incentive compatible if and only if, for all $i$, there exists a constant $\bar{t}_i \in \mathbb{R}$ such that 
\begin{enumerate}
\item For any $i$ and for any $\theta_i \in \Theta_i$, 
\begin{align*}
&\mathbb{E}_{\theta_{-i}}[t_i(\theta_i,\theta_{-i})]\\
=&\bar{t}_i+\mathbb{E}_{\theta_{-i}}\bigg[r_i(\theta)\theta_i\left(\int_V\bm{\mu}_i(\mathbf{v}|\theta)F(\diff \mathbf{v})+\kappa_i\right)+\int_{\theta_i}^{\overline{\theta}_i}r_i(x,\theta_{-i})\bigg(\int_V \bm{\mu}_i(\mathbf{v}|x,\theta_{-i})F(\diff \mathbf{v})+\kappa_i\bigg)\diff x\bigg].
\end{align*}
\item For any $i$, the function 
\[
\theta_i \mapsto \mathbb{E}_{\theta_{-i}}\left[r_i(\theta_i,\theta_{-i})\left(\int_V\bm{\mu}_i(\mathbf{v}|\theta_i,\theta_{-i})F(\diff \mathbf{v})+\kappa_i\right)\right]
\]
is nonincreasing. 
\end{enumerate}
\end{lem}

From \bref{rev}, for any incentive compatible mechanism $(r,\bm{\mu},t)$, and for all $i$, 

The expected consumer surplus can be written as
\begin{align*}
\Sigma(r,\bm{\mu},t)
=& \mathbb{E}_{\theta}\left[\sum_{i=1}^Nr_i(\theta_i,\theta_{-i})\int_V v_i\bm{\mu}_i(\mathbf{v}|\theta_i,\theta_{-i})F(\diff \mathbf{v})\right]\\
&-\sum_{i=1}^N\int_{\Theta_i} \theta_i\mathbb{E}_{\theta_{-i}}\left[r_i(\theta_i,\theta_{-i})\left(\int_V\bm{\mu}_i(\mathbf{v}|\theta)F(\diff \mathbf{v})+\kappa_i\right)\right]G_i(\diff \theta_i)\\
&-\sum_{i=1}^N\int_{\Theta_i}G_i(\theta_i)r_i(\theta_i,\theta_{-i})\mathbb{E}_{\theta_{-i}}\left[r_i(\theta_i,\theta_{-i})\left(\int_V\bm{\mu}_i(\mathbf{v}|\theta_i,\theta_{-i})F(\diff \mathbf{v})+\kappa_i\right)\right]\diff \theta_i-\sum_{i=1}^N \bar{t}_i.
\end{align*}
Meanwhile, for each firm $i$,
\begin{align*}
\int_{\Theta_i}\Pi_i(\theta_i|r,\bm{\mu},t)\Lambda_i(\diff \theta_i)=\int_{\Theta_i}\Lambda_i(\theta_i)\mathbb{E}_{\theta_{-i}}\left[r_i(\theta_i,\theta_{-i})\left(\int_V\bm{\mu}_i(\mathbf{v}|\theta_i,\theta_{-i})F(\diff \mathbf{v})+\kappa_i\right)\right]\diff \theta_i+\Lambda_i(\overline{\theta}_i)\bar{t_i}.
\end{align*}

With the above expressions, we now consider a relaxed problem of \eqref{objj}. To this end, we first introduce the following lemma summarizing the virtual cost functions $\{\phi_i^{\Lambda_i}\}$.
\begin{lem}\label{vc}
For any $i$ and for any nondecreasing, right-continuous function $\Lambda_i$ with $0\leq \Lambda_i(\theta_i)\leq G_i(\theta_i)$, there exists a nondecreasing function $\phi_i^{\Lambda_i}:\Theta_i \to \mathbb{R}_+$ such that
\[
\int_{\Theta_i}\theta_iQ_i(\theta_i)G(\diff \theta_i)+\int_{\Theta_i}(G_i(\theta_i)-\Lambda_i(\theta_i))Q_i(\theta_i)\diff \theta_i \geq \int_{\Theta_i} \phi_i^{\Lambda_i}(\theta_i)Q_i(\theta_i)G_i(\diff \theta_i)
\]
for any nonincreasing function $Q_i:\Theta_i \to \mathbb{R}_+$, and the equality holds whenever $Q_i$ is measurable with respect to the $\sigma$-algebra generated by $\phi_i^{\Lambda_i}$.
\end{lem}

\bref{vc} is essentially the ``ironing'' technique \'{a} la \citet{M81}, except that (i) the type distribution does not necessarily have a density, and (ii) the function being ``ironed'' is the Pareto-weight-adjusted virtual cost, rather than the virtual value. The proof of the lemma follows from \citet*{MS10}, who provide an extension (to even more general settings than ours) of the Myersonian ironing technique that can accommodate these two differences.\footnote{To better understand this lemma, consider the special case when $G_i$ has a density $g_i$ and that $g_i(\theta_i)>0$ for all $\theta_i \in \Theta_i$. Let $\psi_i(\theta_i):=\theta_i+(G_i(
\theta_i)-\Lambda_i(\theta_i))/g_i(\theta_i)$ for all $\theta_i \in \Theta_i$. Then, for any (measurable) function $Q_i:\Theta_i \to \mathbb{R}_+$,
\begin{align*}
\int_{\Theta_i}\theta_iQ_i(\theta_i)G_i(\diff \theta_i)+\int_{\Theta_i}(G_i(\theta_i)-\Lambda_i(\theta_i))Q_i(\theta_i)\diff \theta_i=& \int_{\Theta_i} Q_i(\theta_i)\left(\theta_i+\frac{G_i(\theta_i)-\Lambda_i(\theta_i))}{g_i(\theta_i)}\right)G_i(\diff \theta_i)\\
=&\int_{\Theta_i} Q_i(\theta_i)\psi_i(\theta_i) G_i(\theta_i).
\end{align*}
The function $\psi_i$ is the usual virtual cost function, but since the Pareto weight $\Lambda_i$ is arbitrary, $\psi_i$ is not necessarily monotone. Applying the standard ironing technique to $\psi_i$, we would then obtain the (ironed) virtual cost $\phi_i^{\Lambda_i}$ as described in \bref{vc}.}

Combining \bref{rev} and \bref{vc}, one can observe that the value of \eqref{objj} is bounded from above by the solution of 
\begin{equation}\label{opt}
\sup_{r,\bm{\mu}}\left\{\mathbb{E}_\theta\left[\sum_{i=1}^N r_i(\theta)\left(\int_V \left(v_i-\phi_i^{\Lambda_i}(\theta_i)\right)\bm{\mu}_i(\mathbf{v}|\theta)F(\diff \mathbf{v})-\phi_i^{\Lambda_i}(\theta_i)\kappa_i\right)\right]-\sum_{i=1}^N(1-\Lambda_i(\overline{\theta}_i))\bar{t}_i\right\},
\end{equation}
subject to 
\begin{equation}\label{mon}
\theta_i \mapsto \mathbb{E}_{\theta_{-i}}\left[r_i(\theta_i,\theta_{-i})\left(\int_V\bm{\mu}_i(\mathbf{v}|\theta_i,\theta_{-i})F(\diff \mathbf{v})+\kappa_i\right)\right] \mbox{ is nonincreasing.}
\end{equation}
Moreover, by \bref{rev}, any individually rational mechanism must have $\bar{t}_i \geq 0$ for all $i$. Thus, it is without loss to set $\bar{t}_i=0$ for all $i$. 

In what follows, we characterize the solution of \eqref{objj} by finding a solution to \eqref{opt} first and then verifying that the objective of \eqref{objj} equals the objective of \eqref{opt} under this solution.  To this end, define  $(r^*,\bm{\mu}^*,t^*)$ as follows: For any $\theta \in \Theta$, let $\mathcal{E}^*(\theta)$ be a solution of 
\[
\max_{\mathcal{E}\subseteq \{1,\ldots,N\}}\left(\int_V \max_{i \in \mathcal{E}}(v_i-\phi_i^{\Lambda_i}(\theta_i))^+F(\diff \mathbf{v})-\sum_{i \in \mathcal{E}}\phi_i^{\Lambda}(\theta_i)\kappa_i\right).
\]
Then, let
\[
\bm{\mu}^*_i(\mathbf{v}|\theta):=\left\{
\begin{array}{cc}
\frac{1}{|\mathbb{M}^*(\mathbf{v},\theta)|},     &\mbox{if }v_i \geq \phi_i^{\Lambda_i}(\theta_i)\mbox{ and } i \in \mathbb{M}^*(\mathbf{v},\theta)    \\
0,     &\mbox{otherwise} 
\end{array}
\right.,
\]
where $\mathbb{M}^*(\mathbf{v},\theta):=\argmax_{j \in \mathcal{E}^*(\theta)} \{v_j-\phi_j^{\Lambda_j}(\theta_j)\}$, for all $i$, for all $\mathbf{v} \in V$, and for all $\theta \in \Theta$; and
\[
r^*_i(\theta)=\mathbf{1}\{i \in \mathcal{E}^*(\theta)\}
\]
for all $i$ and for all $\theta \in \Theta$; 
and 
\begin{align*}
&t_i^*(\theta)=T_i^*(\theta_i)\\
:=&\mathbb{E}_{\theta_{-i}}\bigg[r_i^*(\theta)\theta_i\left(\int_V\bm{\mu}^*_i(\mathbf{v}|\theta)F(\diff \mathbf{v})+\kappa_i\right)-\int_{\theta_i}^{\overline{\theta}_i}r_i^*(x,\theta_{-i})\bigg(\int_V \bm{\mu}^*_i(\mathbf{v}|x,\theta_{-i})F(\diff \mathbf{v})+\kappa_i\bigg)\diff x\bigg],
\end{align*}
for all $i$ and for all $\theta \in \Theta$. 

\begin{lem}\label{opm}
The mechanism $(r^*,\bm{\mu}^*,t^*)$ solves \eqref{opt}. Furthermore, 
\begin{align*}
&\sum_{i=1}^N \int_{\Theta_i}\Pi(\theta_i|r^*,\bm{\mu}^*,t^*)\Lambda_i(\diff \theta_i)+\Sigma(r^*,\bm{\mu}^*,t^*)\\
=&\mathbb{E}_{\theta}\left[\sum_{i=1}^N r_i^*(\theta)\left(\int_V (v_i-\phi_i^{\Lambda_i}(\theta_i))\bm{\mu}_i^*(\mathbf{v}|\theta)F(\diff \mathbf{v})-\phi_i^{\Lambda_i}(\theta_i)\kappa_i\right)\right].
\end{align*}
\end{lem}


\bref{opm} implies that the mechanism $(r^*,\bm{\mu}^*,t^*)$ is a solution to \eqref{objj}. Furthermore, \bref{rev} and \bref{vc} imply that any other solution of \eqref{objj} must be outcome-equivalent to $(r^*,\bm{\mu}^*,t^*)$ with probability 1, save for the tie breaking rules that do not affect efficiency.

Now consider any efficient mechanism. As noted above, it is without loss to assume that this mechanism is $(r^*,\bm{\mu}^*,t^*)$. To see that $(r^*,\bm{\mu}^*,t^*)$ is equivalent to a PRYCE CAP mechanism, consider the mechanism $(S,r^{\mathcal{P}},\bm{\mu}_i^{\mathcal{P}},t^{\mathcal{P}})$ as follows: $S_i:=\mathbb{R}_+$ for all $i$; 
\[
r_i^{\mathcal{P}}(s):=\mathbf{1}\{ i \in \mathcal{E}^{\mathcal{P}}(s)\},
\]
for all $s \in S$, where $\mathcal{E}^{\mathcal{P}}(s)$ is a solution of 
\[
\max_{\mathcal{E}\subseteq \{1,\ldots,N\}} \left(\int_V \max_{i \in \mathcal{E}}(v_i-s_i)^+F(\diff \mathbf{v})-\sum_{i \in \mathcal{E}}s_i\kappa_i\right),
\]
for all $s \in S$;
\[
\bm{\mu}_i^{\mathcal{P}}(\mathbf{v}|s)=\left\{
\begin{array}{cc}
\frac{1}{|\mathbb{M}(\mathbf{v},s)|},     &\mbox{if } v_i \geq s_i\mbox{ and } i \in \mathbb{M}(\mathbf{v},s)\\
0,     &\mbox{otherwise} 
\end{array}
\right.,
\]
where  $\mathbb{M}(\mathbf{v},s):=\argmax_{j \in \mathcal{E}^{\mathcal{P}}(s)} \{v_j-s_j\}$; and 
\[
t_i^{\mathcal{P}}(s):=s_i\mathbb{E}_{\theta_{-i}}\left[r_i^{\mathcal{P}}(s_i,\phi_{-i}^{\Lambda_{-i}}(\theta_{-i}))\int_V\bm{\mu}_i^{\mathcal{P}}(\mathbf{v}|s_i,\phi_{-i}^{\Lambda_{-i}}(\theta_{-i}))F(\diff \mathbf{v})\right]-\tau_i^*((\phi_i^{\Lambda_i})^{-1}(s_i)),
\]
where 
\[
\tau_i^*(\theta_i):=\phi_i^{\Lambda_i}(\theta_i)\mathbb{E}_{\theta_{-i}}\left[r_i^*(\theta)\int_V\bm{\mu}^*_i(\mathbf{v}|\theta)F(\diff \mathbf{v})\right]-T_i^*(\theta_i),
\]
and $(\phi_{i}^{\Lambda_{i}})^{-1}(s_i):=\inf\{\theta_i \in \Theta_i|\phi_i^{\Lambda_i}(\theta_i) \geq s_i\}$, for all $i$ and for all $s_i \in S_i$. 

Notice that for any $i$, any $s_{-i} \in S_{-i}$, and any $s_i,s_i' \in S_i$, if $s_i>s_i'$ and $i \in \mathcal{E}^{\mathcal{P}}(s_i,s_{-i})$, it must be that $i \in \mathcal{E}^{\mathcal{P}}(s_i',s_{-i})$. Thus, for any $i$ and for any $s_{-i} \in S_{-i}$, there exists $\bar{p}_i(s_{-i}) \in \mathbb{R}_+ \cup \{\infty\}$ such that $i \in \mathcal{E}^{\mathcal{P}}(s_i,s_{-i})$ if and only if $s_i \leq \bar{p}_i(s_{-i})$. For this reason, $(S,r^{\mathcal{P}},\bm{\mu}^{\mathcal{P}},t^{\mathcal{P}})$ is indeed a PRYCE CAP mechanism.

The following lemma completes the proof.

\begin{lem}\label{lem3}
The PRYCE CAP mechanism $(S,r^{\mathcal{P}},\bm{\mu}_i^{\mathcal{P}},t^{\mathcal{P}})$ has a pure-strategy Bayes-Nash equilibrium $\sigma^{\mathcal{P}}$ that induces the same outcome as $(r^*,\bm{\mu}^*,t^*)$.
\end{lem}

\section{PRYCE CAP Example and Properties} 
\label{sec:example_and_properties}
\subsection{PRYCE CAP Example}
\label{sec:example}

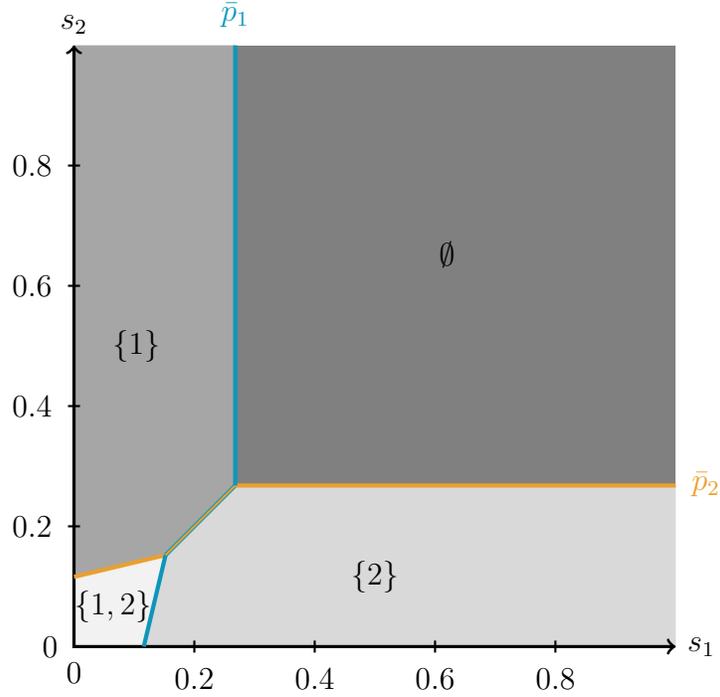
\begin{figure}
    \centering
    \begin{tikzpicture}[scale=8]
\fill [gray!10] (0,0)--(0.11597,0)--(0.15220,0.15220)--(0,0.11597);
\fill [gray!30] (0.11597,0)--(0.15220,0.15220)--(0.26795,0.26795)--(1,0.26795)--(1,0);
\fill [gray!70] (0,0.11597)--(0.15220,0.15220)--(0.26795,0.26795)--(0.26795,1)--(0,1);
\fill [gray!100] (0.26795,0.26795) rectangle (1,1);      
\draw [<->, very thick] (0,1) node (yaxis) [above] {$s_2$}
        |- (1,0) node (xaxis) [right] {$s_1$};
\draw [thick] (0.2,-0.01)--(0.2,0.01); 
\draw (0.2,-0.01) node [below=2pt] {$0.2$};
\draw [thick] (0.4,-0.01)--(0.4,0.01); 
\draw (0.4,-0.01) node [below=2pt] {$0.4$};
\draw [thick] (0.6,-0.01)--(0.6,0.01); 
\draw (0.6,-0.01) node [below=2pt] {$0.6$};
\draw [thick] (0.8,-0.01)--(0.8,0.01); 
\draw (0.8,-0.01) node [below=2pt] {$0.8$};
\draw [thick] (-0.01,0.2)--(0.01,0.2); 
\draw (-0.01,0.2) node [left=2pt] {$0.2$};
\draw [thick] (-0.01,0.4)--(0.01,0.4); 
\draw (-0.01,0.4) node [left=2pt] {$0.4$};
\draw [thick] (-0.01,0.6)--(0.01,0.6); 
\draw (-0.01,0.6) node [left=2pt] {$0.6$};
\draw [thick] (-0.01,0.8)--(0.01,0.8); 
\draw (-0.01,0.8) node [left=2pt] {$0.8$};
\draw (0,0) node [below=2pt] {$0$};
\draw (0,0) node [left=2pt] {$0$};

\draw [GoldOrangeT, ultra thick] (0,0.11597)--(0.15220,0.15220);
\draw [BlueGreenT, ultra thick] (0.11597,0)--(0.15220,0.15220);
\draw [BlueGreenT, ultra thick] (0.15220,0.15220)--(0.26795,0.26795);
\draw [GoldOrangeT, thick] (0.15220,0.15220)--(0.26795,0.26795);
\draw [GoldOrangeT, ultra thick] (0.26795,0.26795)--(1,0.26795);
\draw [BlueGreenT, ultra thick] (0.26795,0.26795)--(0.26795,1);
\draw (0.0645,0.12) node [below=2pt] {$\{1,2\}$};
\draw (0.5,0.17) node [below=2pt] {$\{2\}$};
\draw (0.17,0.5) node [left=2pt] {$\{1\}$};
\draw (0.62,0.7) node [below=2pt] {$\emptyset$};

\draw (1.05,0.27) node {\color{GoldOrangeT}$\bar{p}_2$};
\draw (0.26795,1.05) node {\color{BlueGreenT}$\bar{p}_1$};

\end{tikzpicture}
\caption{Values of $\mathcal{E}^{\mathcal{P}}$}
\label{fig1}
\end{figure}

Suppose that the number of potentially active firms $N=2$ and that consumer values and firm types $v_1, \theta_1,v_2,\theta_2 \in [0,1]$ are independently drawn from a uniform distribution. Suppose further that the commonly known fixed cost parameters $\kappa_1=\kappa_2=1$ and that the Pareto weight functions $\Lambda_1(x)=\Lambda_2(x)=x$ for all $x \in [0,1]$. Then, the price cap functions $\bar{p}_1$ and $\bar{p}_2$ and the set $\mathcal{E}^{\mathcal{P}}$ can be depicted by \bref{fig1}. 

\bref{fig1} illustrates the tight link between the yardstick price caps and the sets of firms optimally granted market entry. In the figure, the set of (undominated) prices $[0,1]^2$ is partitioned into four regions, where each region of $(s_1,s_2)$ is mapped into different values of $\mathcal{E}^{\mathcal{P}}(s_1,s_2)$. As a result, the boundaries of the regions define the yardstick price caps. The gold-orange curve represents firm 2's price cap $\bar{p}_2$ as a function of $s_1$, and the blue-green curve represents firm 1's price cap $\bar{p}_1$ as a function of $s_2$. 

Given firm 1's published price $s_1$, firm 2 is excluded from the market if it posts a price $s_2>\bar{p}_2(s_1)$. Similarly, given firm 2's published price $s_2$, firm 1 is excluded from the market if it posts a price $s_1>\bar{p}_1(s_2)$. Notice that both firms operate in the market if both publish relatively low prices, and both are restricted from entering if both publish relatively high prices. If one firm posts too high a price relative to the second, the first firm is excluded, whereas the second can enter.

Focusing on the behavior of the price caps in the figure, one can observe the two caps initially increasing in the other firm's price. As the competing firm publishes a higher price, the restriction on the other firm's price loosens, consistent with the yardstick nature of the price ceiling. Once the competing firm's price exceeds a certain value, though, the other firm's price cap flattens, becoming independent of the competing firm's choice. This change in pattern is from the competing firm no longer operating in the market precisely because its high price denied it entry. At that point, the price cap of the other firm remains fixed and its authorization for business depends only on its own published price.

\subsection{Yardstick Price Cap Properties}
\label{sec:properties}
The properties of the yardstick price cap just described are not special to the assumptions of two firms or uniformly distributed consumer values. Under the broader assumption that consumers' values $\{v_i\}_{i=1}^N$ are i.i.d., the next proposition explains that a firm's price ceiling rises when competing firms submit higher prices. Moreover, a firm can guarantee itself entry if it submits a price \textit{below} a certain threshold; and it can guarantee itself no entry if it submits a price \textit{above} another threshold. 

\begin{prop}\label{prop:property}
Suppose that $\{v_i\}_{i=1}^N$ are i.i.d. and that $\kappa_i=\kappa$ for all $i$. Consider any efficient PRYCE CAP mechanism and let $\bar{p}_i:S_{-i} \to \mathbb{R}_+ \cup \{\infty\}$ denote the yardstick price cap for firm $i$. Then, 
\begin{enumerate}
\item For any price vector $s \in \mathbb{R}_+^N$, $\bar{p}_i(s_{-i}) \leq \bar{p}_j(s_{-j})$ if and only if $s_i \geq s_j$, for all $i,j \in \{1,\ldots,N\}$.
\item For any $i \in \{1,\ldots,N\}$ and for any $s_{-i} \in \mathbb{R}_+^{N-1}$, $\bar{p}_i(s_{-i}) \in [\underline{s},\bar{s}]$ for some $0\leq \underline{s} \leq \bar{s}<\infty$.  
\end{enumerate}
\end{prop}

An immediate consequence of \bref{prop:property} is that the firm publishing the lowest price faces the highest price ceiling. This relation implies that the price a firm publishes has two effects on its eligibility to operate in an efficient PRYCE CAP mechanism. The first is a \emph{direct effect}: A lower submitted price is more likely to be below the firm's price ceiling and grant the firm the right to sell. The second is a \emph{yardstick effect}: A lower submitted price, other things equal, means the firm will face a higher price ceiling compared to its competitors, which can be more easily met. 

\section{Extensions}
\label{sec:extensions}
\subsection {Efficient Mechanism under Capacity Constraints}
The arguments in the baseline model can be readily extended to a setting where firms are subject to (publicly known) capacity constraints. Specifically, suppose that, in addition to facing a cost function $C_i$, each firm $i$ faces a capacity constraint $L_i \geq 0$ such that firm $i$ cannot produce more than $L_i \geq 0$ units. Namely, for any incentive compatible and individually rational direct mechanism $(r,\bm{\mu},t)$, and for all $i$,
\begin{equation}\label{capacity}
\int_V \bm{\mu}_i(\mathbf{v}|\theta)F(\diff \mathbf{v}) \leq L_i.
\end{equation}

Consequently, the maximization problem that characterizes efficient mechanisms would simply be \eqref{objj}, but subject to \eqref{capacity}, and \eqref{ic}, \eqref{ir}. The additional capacity constraints can be incorporated into the arguments above via the duality theorem. Specifically, let $\lambda_i \geq 0$ be the Lagrange multiplier of the capacity constraint \eqref{capacity} for firm $i$. The Lagrangian to the relaxed problem \eqref{opt} is simply 
\[
\mathbb{E}_\theta\left[\sum_{i=1}^N r_i(\theta)\left(\int_V \left(v_i-\phi_i^{\Lambda_i}(\theta_i)-\lambda_i\right)\bm{\mu}_i(\mathbf{v}|\theta)F(\diff \mathbf{v})-\phi_i^{\Lambda_i}(\theta_i)\kappa_i\right)\right]-\sum_{i=1}^N(1-\Lambda_i(\overline{\theta}_i))\bar{t}_i.
\]

As a result, since $\phi^{\Lambda_i}(\theta_i)-\lambda_i$ remains nondecreasing in $\theta_i$ for each $i$, the arguments above can still be applied even when there are additional capacity constraints. Thus, by properly adjusting the transfers, the efficient mechanism can still be implemented by a PRYCE CAP mechanism even when there are capacity constraints.

\subsection{Consumers' Ex-Post Individual Rationality Constraints}
In the baseline model, monetary transfers are captured entirely by $\{t_i\}_{i=1}^N$. In particular, there is no distinction between prices that consumers pay to firms, taxes that firms pay to a regulator, and subsidies the regulator pays to consumers. While convenient, this modeling approach cannot incorporate consumers' ex-post individual rationality constraints, and consumers might be forced to purchase a product even if the price is higher than their values under some indirect mechanisms. Nonetheless, the results above can be easily extended to a setting where consumers' ex-post individual rationality constraints are taken into account. 

To see this, consider an alternative formulation of indirect mechanism $(S_i,r_i,\bm{\mu}_i,\bm{p}_i,\tau_i)_{i=1}^N$, where $S_i$, $r_i$, and $\bm{\mu}_i$ are the same as defined in the baseline model, while $\bm{p}_i(s)$ denotes firm $i$'s transaction price when the strategy profile is $s$; and $\tau_i(s)$ denotes the amount of (lump-sum) transfers paid by firm $i$ when the strategy profile is $s$. Consumers' ex-post individual rationality constraints are given by 
\begin{equation}\label{epir}
\sum_{i=1}^N\max\{(\bm{p}_i(s))-v_i),0\}\bm{\mu}_i(\mathbf{v}|s)=0,
\end{equation}
for all $s \in S$ and for all $\mathbf{v} \in V$. In other words, an indirect mechanism is ex-post individually rational if, for any strategy profiles adopted by firms and for any realized consumer values, a consumer with value $\mathbf{v}$ buys from firm $i$ only if firm $i$'s price $\bm{p}_i(s)$ is lower than the consumer's value $v_i$. 

Note that for any indirect mechanism $(S_i,r_i,\bm{\mu}_i,\bm{p}_i,\tau_i)_{i=1}^N$ that is ex-post individually rational for consumers, let 
\[
t_i(s):=\tau_i(s)-\bm{p}_i(s)\int_V\bm{\mu}_i(\mathbf{v}|s)F(\diff \mathbf{v}),
\]
for all $s \in S$. The mechanism $(S_i,r_i,\bm{\mu}_i,t_i)$ then corresponds to a feasible indirect mechanism in the baseline model. Moreover, any PRYCE CAP mechanism, by definition, is a mechanism where consumers' ex-post individually rational constraint \eqref{epir} is satisfied. Therefore, PRYCE CAP mechanisms remain optimal for the regulator even if consumers are required to be ex-post individually rational. 

\section{Conclusion}\label{sec:conclusion}

We study the optimal regulation of oligopolistic competition where firms have private information about their costs and consumers make discrete choices over goods. We search over a broad class of mechanisms, covering a variety of ways in which firms compete, that implement constrained Pareto efficient allocations. The socially efficient mechanisms are equivalent to price competition, but with lump-sum transfers and yardstick price caps. We refer to these mechanisms as PRYCE CAP mechanisms, and they can be implemented without knowledge of individual consumer preferences, realized firm costs, or firm conduct.  

To implement a PRYCE CAP mechanism, a regulator, we presume, has power to verify and enforce competition exclusively on price, regardless of the kinds of complicated competitive conduct that might already prevail in the market. But upholding price competition is not the unique way to achieve efficiency. For certain markets, a clever selection of lump-sum transfers alone might convert an existing inefficient market setting into an efficient one. But administering such creative transfers would likely be intractable, and the regulator would need unearthly knowledge of the competitive game that firms engage. A significant contribution of PRYCE CAP mechanisms is that they require no such awareness, and they apply to a broad range of potential market environments.

Hence, if the regulator can verify and enforce price competition, the search for efficiency ends with PRYCE CAP mechanisms. In practice, though, a regulator might lack such powers entirely or wield them imperfectly. A natural implication of our result is that social efficiency is more easily achieved in markets where a regulator can plausibly maintain price competition. Or rather, more realistically, markets where posting prices \textit{already} drives the nature of competition are better candidates for reaching efficiency.

\clearpage{}

\begin{spacing}{1.0}

\appendix
\noindent\textbf{\LARGE{Appendix}}

\section{Omitted Proofs for Section \ref{sec:proof_thm1}}
\label{sec:proofs}
\subsection{Proof of Lemma \ref{vc}}
\label{aa1}
Let $\nu_i$ be a signed measure on $\Theta_i:=[\underline{\theta}_i,\overline{\theta}_i]$ defined by 
\[
\nu(A):=\int_A \theta_i G_i(\diff \theta_i)+\int_A (G_i(\theta_i)-\Lambda_i(\theta_i))\diff \theta_i,
\]
for all (Borel) subset $A$ of $\Theta_i$, and let $N_i(\theta_i):=\nu([\underline{\theta}_i,\theta_i])$ for all $\theta_i \in \Theta_i$ be its CDF. By definition 3 and theorem 2 of \citet*{MS10}, there exists a nondecrasing function $\phi_i^{\Lambda_i}$ such that for any nonincreasing function $Q_i:\Theta_i \to \mathbb{R}_+$,
\begin{align*}
&\int_{\Theta_i}\theta_iG_i(\theta_i)(-Q_i(\theta_i))G_i(\diff \theta_i)+\int_{\Theta_i}(G_i(\theta_i)-\Lambda_i(\theta_i))(-Q_i(\theta_i))\diff \theta_i\\
=&\int_{\Theta_i}(-Q_i(\theta_i))\nu_i(\diff \theta_i)\\
\leq& \int_{\Theta_i}(-Q_i(\theta_i))\phi_i^{\Lambda_i}(\theta_i)G_i(\diff \theta_i),
\end{align*}
and hence 
\[
\int_{\Theta_i}\theta_iG_i(\theta_i)Q_i(\theta_i)G_i(\diff \theta_i)+\int_{\Theta_i}(G_i(\theta_i)-\Lambda_i(\theta_i))Q_i(\theta_i)\diff \theta_i \geq \int_{\Theta_i}Q_i(\theta_i)\phi_i^{\Lambda_i}(\theta_i)G_i(\diff \theta_i)
\]
for any nonincreasing function $Q_i$. Meanwhile, theorem 3 of \citet*{MS10} implies that the inequality is binding whenever $Q_i$ is measurable with respect to the $\sigma$-algebra generated by $\phi_i^{\Lambda_i}$. This completes the proof. \hfill $\blacksquare$

\subsection{Proof of Lemma \ref{opm}}
\label{aa3}
\begin{proof}
We first show that, for all $i$,  
\begin{equation}\label{allocation}
\theta_i \mapsto  \mathbb{E}_{\theta_{-i}}\left[r_i^*(\theta_i,\theta_{-i})\left(\int_V \bm{\mu}^*_i(\mathbf{v}|\theta_i,\theta_{-i})F(\diff \mathbf{v})+\kappa_i\right)\right]
\end{equation}
is nonincreasing. To see this, notice that for any $i$ and for any $\theta \in \Theta$,
\begin{equation}\label{q1}
\int_V\bm{\mu}_i^*(\mathbf{v}|\theta)F(\diff \mathbf{v})=\int_V \mathbf{1}\{\phi_i^{\Lambda_i}(\theta_i) \leq \phi_i^{\Lambda_i}(\theta_j)+v_i-v_j, \, \forall j \in \mathcal{E}^*(\theta), \, j \neq i\}F(\diff \mathbf{v}).
\end{equation}
Moreover, notice that for any $i$, for any $\theta_{-i} \in \Theta_{-i}$, and for any $\theta_i,\theta_i' \in \Theta_i$ with $\theta_i'<\theta_i$, $i \in \mathcal{E}^*(\theta_i,\theta_{-i})$ implies $i \in \mathcal{E}^*(\theta_i',\theta_{-i})$. Together with the fact that $\phi_i^{\Lambda_i}$ is nondecreasing, it then follows that both \eqref{q1} and $r_i^*$ are nonincreasing functions of $\theta_i$ for all $\theta_{-i} \in \Theta_{-i}$. Therefore, \eqref{allocation} is indeed nonincreasing. 

Furthermore, by definition of $(r^*,\bm{\mu}^*)$, for any $(r,\bm{\mu})$ such that the function \[\theta_i \mapsto  \mathbb{E}_{\theta_{-i}}\left[r_i(\theta_i,\theta_{-i})\left(\int_V \bm{\mu}_i(\mathbf{v}|\theta_i,\theta_{-i})F(\diff \mathbf{v})+\kappa_i\right)\right]\] is nonincreasing, it must be that 
\begin{align*}
&\mathbb{E}_\theta \left[\sum_{i=1}^N r_i(\theta)\left(\int_V(v_i-\phi_i^{\Lambda_i}(\theta_i))\bm{\mu}_i(\mathbf{v}|\theta)F(\diff \mathbf{v})-\phi_i^{\Lambda_i}(\theta_i)\kappa_i\right)\right]\\
\leq& \mathbb{E}_\theta \left[\sum_{i=1}^N r_i^*(\theta)\left(\int_V(v_i-\phi_i^{\Lambda_i}(\theta_i))\bm{\mu}^*_i(\mathbf{v}|\theta)F(\diff \mathbf{v})-\phi_i^{\Lambda_i}(\theta_i)\kappa_i\right)\right].
\end{align*}
Thus, $(r^*,\bm{\mu}^*)$ is a solution to \eqref{opt}. 

Lastly, by \bref{vc}, since \eqref{allocation} is nonincreasing and is measurable with respect to $\phi_i^{\Lambda_i}$ for all $i$, we have
\begin{align*}
&\int_{\Theta_i}\phi_{i}^{\Lambda_i}(\theta_i)\mathbb{E}_{\theta_{-i}}\left[r_i^*(\theta)\left(\int_V \bm{\mu}_i^*(\mathbf{v}|\theta)F(\diff \mathbf{v})+\kappa_i\right)\right]G_i(\diff \theta_i)\\
=&\int_{\Theta_i}\theta_i\mathbb{E}_{\theta_{-i}}\left[r_i^*(\theta)\left(\int_V \bm{\mu}_i^*(\mathbf{v}|\theta)F(\diff \mathbf{v})+\kappa_i\right)\right]G_i(\diff \theta_i)\\
&+\int_{\Theta_i}(G_i(\theta_i)-\Lambda_i(\theta_i))\mathbb{E}_{\theta_{-i}}\left[r_i^*(\theta)\left(\int_V \bm{\mu}_i^*(\mathbf{v}|\theta)F(\diff \mathbf{v})+\kappa_i\right)\right]\diff \theta_i
\end{align*}

Therefore, 
\begin{align*}
&\mathbb{E}_\theta \left[\sum_{i=1}^N r_i^*(\theta)\left(\int_V(v_i-\phi_i^{\Lambda_i}(\theta_i))\bm{\mu}^*_i(\mathbf{v}|\theta)F(\diff \mathbf{v})-\phi_i^{\Lambda_i}(\theta_i)\kappa_i\right)\right]\\
=&\mathbb{E}_\theta\left[\sum_{i=1}^Nr_i^*(\theta)\int_Vv_i\bm{\mu}_i^*(\mathbf{v}|\theta)F(\diff \mathbf{v})\right]
-\sum_{i=1}^N\left\{\int_{\Theta_i}\phi_{i}^{\Lambda_i}(\theta_i)\mathbb{E}_{\theta_{-i}}\left[r_i^*(\theta)\left(\int_V \bm{\mu}_i^*(\mathbf{v}|\theta)F(\diff \mathbf{v})+\kappa_i\right)\right]G_i(\diff \theta_i)\right\}\\
=&\mathbb{E}_\theta\left[\sum_{i=1}^Nr_i^*(\theta)\int_Vv_i\bm{\mu}_i^*(\mathbf{v}|\theta)F(\diff \mathbf{v})\right]-\int_{\Theta_i}\theta_i\mathbb{E}_{\theta_{-i}}\left[r_i^*(\theta)\left(\int_V \bm{\mu}_i^*(\mathbf{v}|\theta)F(\diff \mathbf{v})+\kappa_i\right)\right]G_i(\diff \theta_i)\\
&-\int_{\Theta_i}(G_i(\theta_i)-\Lambda_i(\theta_i))\mathbb{E}_{\theta_{-i}}\left[r_i^*(\theta)\left(\int_V \bm{\mu}_i^*(\mathbf{v}|\theta)F(\diff \mathbf{v})+\kappa_i\right)\right]\diff \theta_i\\
=&\Sigma(r^*,\bm{\mu}^*,t^*)+\sum_{i=1}^N\int_{\Theta_i}\Pi(\theta_i|r^*,\bm{\mu^*},t^*)\Lambda_i(\diff \theta_i),
\end{align*}
as desired.
\end{proof}

\subsection{Proof of Lemma \ref{lem3}}\label{aa4}
\begin{proof}
Consider the mechanism $(S,r^{\mathcal{P}},\bm{\mu}_i^{\mathcal{P}},t^{\mathcal{P}})$. First, notice that by definition of $\bm{\mu}_i^{\mathcal{P}}$, for all $\theta \in \Theta$ and for all $i$,
\[
\bm{\mu}_i^{\mathcal{P}}(\mathbf{v}|\phi_1^{\Lambda_1}(\theta_1),\ldots,\phi_N^{\Lambda_N}(\theta_N))=\bm{\mu}_i^*(\mathbf{v}|\theta_1,\ldots,\theta_N),
\]
for all $\mathbf{v} \in V$. Moreover, by \bref{rev}, for each $i$ and for any interval $[\theta^1_i,\theta^2_i]$ on which $\phi_i^{\Lambda_i}$ is constant, $T_i^*$ is also constant. Therefore, for any $i$ and for any $\theta_i \in \Theta_i$, if $\theta_i$ belongs to an interval $[\theta^1_i,\theta^2_i]$ on which $\phi_i^{\Lambda_i}$ is a constant, then $(\phi_i^{\Lambda_i})^{-1}(\phi_i^{\Lambda_i}(\theta_i))=\theta_i^2=\theta_i$, Thus, for any $i$ and for any $\theta \in \Theta$, 
\begin{align*}
t_i^{\mathcal{P}}(\phi_1^{\Lambda_1}(\theta_1),\ldots,\phi_N^{\Lambda_N}(\theta_N))=&\mathbb{E}_{\theta_{-i}}\left[\phi_i^{\Lambda_i}(\theta_i)r_i^{\mathcal{P}}(\phi_i^{\Lambda_i}(\theta_i),\phi_{-i}^{\Lambda_{-i}}\theta_{-i})\int_V\bm{\mu}_i^{\mathcal{P}}(\mathbf{v}|\phi_i^{\Lambda_i}(\theta_i),\phi_{-i}^{\Lambda_{-i}}(\theta_{-i}))F(\diff \mathbf{v})\right]-\tau_i^*(\theta_i)\\
=& \mathbb{E}_{\theta_{-i}}\left[\phi_i^{\Lambda_i}(\theta_i)r_i^*(\theta)\int_V\bm{\mu}^*_i(\mathbf{v}|\theta)F(\diff \mathbf{v})\right]-\tau_i^*(\theta_i)\\
=&T_i^*(\theta_i)\\
=&t_i^*(\theta_1,\ldots,\theta_N),
\end{align*}
where $\phi_{-i}^{\Lambda_{-i}}:=(\phi_1^{\Lambda_1},\ldots,\phi_{i-1}^{\Lambda_{i-1}},\phi_{i+1}^{\Lambda_{i+1}},\ldots,\phi_N^{\Lambda_N})$. Furthermore, by the definitions of $\mathcal{E}^{\mathcal{P}}$ and $\mathcal{E}^*$ given $\mu^{\mathcal{P}}$ and $t^{\mathcal{P}}$, when each firm $i$ with type $\theta_i$ chooses $\phi_i^{\Lambda_i}(\theta_i)$, the induced welfare outcomes (i.e., the weighted sum of consumer surplus and firms' interim expected revenue) under $(r^*,\mu^*,t^*)$ is the same as that under $(S,r^{\mathcal{P}},\mu^{\mathcal{P}},t^{\mathcal{P}})$. 

It then remains to show that the strategy profile where each firm $i$ with type $\theta_i$ chooses $\phi_i^{\Lambda_i}(\theta_i)$ is a Bayes-Nash equilibrium in the game induced by $(S,r^{\mathcal{P}},\bm{\mu}_i^{\mathcal{P}},t^{\mathcal{P}})$. Indeed, for any firm $i$, any type $\theta_i \in \Theta_i$, and for any $s_i \in \phi_i^{\Lambda_i}(\Theta_i)$, given that all other firms follow the strategy $\phi_{-i}^{\Lambda_{-i}}=(\phi_1^{\Lambda_1},\ldots,\phi_{i-1}^{\Lambda_{i-1}},\phi_{i+1}^{\Lambda_{i+1}},\ldots,\phi_N^{\Lambda_N})$, let $\theta_i' \in \Theta_i$ be such that $\phi_i^{\Lambda_i}(\theta_i')=s_i$. We then have 
\begin{align*}
&\mathbb{E}_{\theta_{-i}}\bigg[t_i^{\mathcal{P}}(\phi_i^{\Lambda_i}(\theta_i),\phi_{-i}^{\Lambda_{-i}}(\theta_{-i}))-r_i^{\mathcal{P}}(\phi_i^{\Lambda_i}(\theta_i),\phi_{-i}^{\Lambda_{-i}}(\theta_{-i}))\theta_i\left(\int_V \bm{\mu}_i^{\mathcal{P}}(\mathbf{v}|\phi_i^{\Lambda_i}(\theta_i),\phi_{-i}^{\Lambda_{-i}}(\theta_{-i}))F(\diff \mathbf{v})+\kappa_i\right)\bigg]\\
=&\mathbb{E}_{\theta_{-i}}\left[t_i^*(\theta_i,\theta_{-i})-r_i^*(\theta_i,\theta_{-i})\theta_i\left(\int_V\bm{\mu}^*_i(\mathbf{v}|\theta_i,\theta_{-i})F(\diff \mathbf{v})+\kappa_i\right)\right] \\
\geq& \mathbb{E}_{\theta_{-i}}\left[t_i^*(\theta_i',\theta_{-i})-r_i^*(\theta_i',\theta_{-i})\theta_i\left(\int_V\bm{\mu}^*_i(\mathbf{v}|\theta_i',\theta_{-i})F(\diff \mathbf{v})+\kappa_i\right)\right]\\
=&\mathbb{E}_{\theta_{-i}}\bigg[t_i^{\mathcal{P}}(\phi_i^{\Lambda_i}(\theta_i'),\phi_{-i}^{\Lambda_{-i}}(\theta_{-i}))-r_i^{\mathcal{P}}(\phi_i^{\Lambda_i}(\theta_i'),\phi_{-i}^{\Lambda_{-i}}(\theta_{-i}))\theta_i\left(\int_V \bm{\mu}_i^{\mathcal{P}}(\mathbf{v}|\phi_i^{\Lambda_i}(\theta_i'),\phi_{-i}^{\Lambda_{-i}}(\theta_{-i}))F(\diff \mathbf{v})+\kappa_i\right)\\
&\mathbb{E}_{\theta_{-i}}\bigg[t_i^{\mathcal{P}}(s_i,\phi_{-i}^{\Lambda_{-i}}(\theta_{-i}))-r_i^{\mathcal{P}}(s_i,\phi_{-i}^{\Lambda_{-i}}(\theta_{-i}))\theta_i\left(\int_V \bm{\mu}_i^{\mathcal{P}}(\mathbf{v}|s_i,\phi_{-i}^{\Lambda_{-i}}(\theta_{-i}))F(\diff \mathbf{v})+\kappa_i\right),
\end{align*}
where the inequality follows from the fact that $(r^*,\bm{\mu}^*,t^*)$ is incentive compatible. Meanwhile, it is easy to verify that  for any firm $i$, any type $\theta_i \in \Theta_i$, and for any $s_i \notin \phi_i^{\Lambda_i}(\Theta_i)$, given that all other firms follow the strategy $\phi_{-i}^{\Lambda_{-i}}$, 
\begin{align*}
&\mathbb{E}_{\theta_{-i}}\bigg[t_i^{\mathcal{P}}(\phi_i^{\Lambda_i}(\theta_i'),\phi_{-i}^{\Lambda_{-i}}(\theta_{-i}))-r_i^{\mathcal{P}}(\phi_i^{\Lambda_i}(\theta_i),\phi_{-i}^{\Lambda_{-i}}(\theta_{-i}))\theta_i\left(\int_V \bm{\mu}_i^{\mathcal{P}}(\mathbf{v}|\phi_i^{\Lambda_i}(\theta_i),\phi_{-i}^{\Lambda_{-i}}(\theta_{-i}))F(\diff \mathbf{v})+\kappa_i\right)\\
\geq &\mathbb{E}_{\theta_{-i}}\bigg[t_i^{\mathcal{P}}(\phi_i^{\Lambda_i}(\theta_i'),\phi_{-i}^{\Lambda_{-i}}(\theta_{-i}))-r_i^{\mathcal{P}}(s_i,\phi_{-i}^{\Lambda_{-i}}(\theta_{-i}))\theta_i\left(\int_V \bm{\mu}_i^{\mathcal{P}}(\mathbf{v}|s_i,\phi_{-i}^{\Lambda_{-i}}(\theta_{-i}))F(\diff \mathbf{v})+\kappa_i\right).
\end{align*}
Together, it then follows that $(\phi_1^{\Lambda_1},\ldots,\phi_N^{\Lambda_N})$ is indeed a Bayes-Nash equilibrium in the game induced by $(S,r^{\mathcal{P}},\bm{\mu}_i^{\mathcal{P}},t^{\mathcal{P}})$. This completes the proof.
\end{proof}

\section{Omitted Proof for Section \ref{sec:example_and_properties}
}

\subsection{Proof of Proposition \ref{prop:property}}\label{ab1}
\begin{proof}
From the proof of \bref{thm1}, for each $i \in \{1,\ldots,N\}$ and for any $s \in \mathbb{R}_+^N$, firm $i \in \mathcal{E}^{\mathcal{P}}(s_i,s_{-i})$ if and only if $s_i \leq \bar{p}_i(s_{-i})$, where $\mathcal{E}^P$ is a solution of 
\[
\max_{\mathcal{E} \subseteq \{1,\ldots,N\}} \left(\int_0^\infty \cdots \int_0^\infty \max_{i \in \mathcal{E}}(v_i-s_i)F(\diff v_1)\cdots F(\diff v_N)-\sum_{i \in \mathcal{E}}s_i\kappa_i\right).
\]
As a result, there must exists $\bar{p}: \mathbb{R}_+^{N-1} \to \mathbb{R}_+ \cup \{\infty\}$ such that $\bar{p}_i(s_{-i})=\bar{p}(s_{-i})$ for all $i$ and for all $s \in \mathbb{R}_+^N$. We claim that $\bar{p}$ is nondecreasing in each argument. Indeed, for any $i$ and for any $s,s' \in \mathbb{R}_+^N$, such that $s_i=s_i'$ and $s_j \leq s_j'$ for some $j \neq i$, if $i \in \mathcal{E}^{\mathcal{P}}(s)$, then it must be that $i \in \mathcal{E}^{\mathcal{P}}(s')$ as well. Therefore, it must be that $\bar{p}(s_{-i})\leq \bar{p}(s_{-i}')$, as desired. Since $\bar{p}$ is nondecreasing in every component, for any $i,j \in \{1,\ldots,N\}$ with $i \neq j$, and for any $s \in \mathbb{R}_+^N$ with $s_i \geq s_j$, it must be that $\bar{p}(s_{-j}) \geq \bar{p}(s_{-i})$, as desired.

Meanwhile, notice that for any $i \in \{1,\ldots,N\}$ and for any $s_{-i} \in \mathbb{R}_+^{N-1}$, if $s_i=0$, then it must be that $i \in \mathcal{E}^{\mathcal{P}}(s_i,s_{-i})$. In contrast, since for any $\mathcal{E} \subseteq \{1,\ldots,N\}$ such that $i \notin \mathcal{E}$, 
\[
\lim_{s_i \to \infty} \sup_{s_{-i} \in \mathbb{R}_+^N}\left[\int_0^\infty \cdots\int_0^\infty [\max_{j \in \mathcal{E}\cup \{i\}}(v_j-s_j)^+-\max_{j \in \mathcal{E}}(v_j-s_j)^+]F(\diff v_1)\cdots F(\diff v_N)-s_i\kappa\right]<0,
\]
there must exist $\bar{s}$ such that $i \notin \mathcal{E}^{P}(s)$ whenever $s_i \leq \bar{s}$, for all $s \in \mathbb{R}_+^N$. This completes the proof. 
\end{proof}

\clearpage
\bibliography{ref}

\clearpage

\noindent\textbf{\LARGE{Online Appendix}}
\section*{Omitted Details for Section \ref{sec:range_of_games}}\label{A1}
Here we provide other example indirect mechanisms and provide further details on quantity competition and entry deterrence.

\begin{exa}[\textbf{Entry Deterrence}]
The following indirect mechanism describes a model where an incumbent (firm 1) can use price to deter entry of potential entrants á la \citet{von1934market} sequential competition. Let firm 1's strategy space be $S_1=\mathbb{R}_+$, and let all other firms' strategy spaces be a tuple that consists of entry decisions and prices as functions of firm 1's price. The entry probability $r_1$ of firm 1 is set to $1$ regardless of the strategy profile, whereas entry probabilities of all other firms $i \neq 1$ are given by their strategies. Lastly, revenues $\{t_i\}_{i=1}^N$ and good allocations $\{\bm{\mu}_i\}_{i=1}^N$ are determined by price competition among all firms that enter.
\end{exa}

\begin{exa}[\textbf{Reverse Auction with Many Buyers}]\label{ex6}
The following indirect mechanism describes a type of reverse auction where firms bid their prices and those with the lowest bid win and sell their goods to all consumers with values above that bid. Pharmacy Benefit Managers (PBMs) are known to employ this kind of auction.\footnote{The PBM holds an auction across drugs that treat a particular medical condition. The drugs of the winning manufacturers (the sellers) with the lowest bids are put on the PBM's formulary for doctors to prescribe and patients (the consumers) to purchase.} Under this mechanism, $S=\mathbb{R}_+^N$, $t_i(s)=s_i\int_V \bm{\mu}_i(\mathbf{v}|s)F(\diff \mathbf{v})$, $r_i(s)=1$ for all $s \in S$ and for all $i$, and
\[
\bm{\mu}_i(\mathbf{v}|s)=\left\{
\begin{array}{cc}
\frac{\mathbf{1}\{s_i \in \mathbb{M}(s)\}}{|\mathbb{M}(s)|},& \mbox{if } s_i \leq v_i\\
0,& \mbox{if } s_i > v_i
\end{array}
\right.,
\]
for all $i$, for all $\mathbf{v} \in V$, and for all $s \in S$, where $\mathbb{M}(s):=\argmin_i\{s_i\}$.
\end{exa}

\noindent \textbf{Details for the Quantity Competition Mechanism: }
Let the strategy space be $S=[0,1]^N$, and let entry probabilities be $r_i(s)=1$ for all $i$ and for all $s \in S$. Furthermore, since $F$ is atomless, the function $(p_1,\ldots,p_N) \mapsto \int_V \mathbf{1}\{v_i \geq p_i, \, \forall i\}F(\diff \mathbf{v})$ is continuous and nondecreasing and has value $1$ at $p_1=\ldots=p_N=0$ and value $0$ at $p_1=\ldots=p_N=\max_i\{\max(\mathrm{proj}_i V)\}$. Therefore, for any $s \in S$ such that $\sum_j s_j \leq 1$, there exists $\{\bm{p}_i(s)\}_{i=1}^N$ such that 
\[
\int_V \mathbf{1}\{v_i \geq \bm{p}_i(s), \, \forall i\}F(\diff \mathbf{v})=\sum_{j=1}^Ns_j.
\]
Take any of such functions, and, for each $i$, extend $\bm{p}_i$ to be defined on the entire $S$ by letting $\bm{p}_i(s)=0$ for all $s \in S$ such that $\sum_j s_j>1$. 

Now let 
\[
\bm{\mu}_i(\mathbf{v}|s):=\left\{
\begin{array}{cc}
\frac{s_i}{\sum_j s_j},& \mbox{if } v_j \geq \bm{p}_i(s), \, \forall j\\
0,&\mbox{otherwise}
\end{array}
\right.,
\]
for all $i$, for all $\mathbf{v} \in V$, and for all $s \in S$, and let the revenues be defined as $t_i(s)=\bm{p}_i(s)\int_V\bm{\mu}_i(\mathbf{v}|s)F(\diff \mathbf{v})$.  \\

\noindent \textbf{Details for the Entry Deterrence Mechanism: }
Let $S_1:=\mathbb{R}_+$ and let $S_i:=\{e_i:S_1 \to \{0,1\}\} \times \{\bm{p}_i:S_1 \to \mathbb{R}_+\}$, so that a strategy $s_i$ for firm $i>1$ can be written as $s_i=(e_i,\bm{p}_i)$. Define the revenues as $t_i(s_1,(e_i,\bm{p}_i)_{i=2}^N)=\bm{p}_i(s_1)\int_V\bm{\mu}_i(\mathbf{v}|s)F(\diff \mathbf{v})$, and let the entry probabilities be $r_1(s_1,(e_i,\bm{p}_i)_{i=2}^N)=1$ and $r_i(s_1,(e_i,\bm{p}_i)_{i=2}^N)=e_i(s_1)$, for all $s \in S$ and for all $i > 1$. Finally, let $\bm{\mu}$ be defined as 
\begin{align*}
&\bm{\mu}_i(\mathbf{v}|s_1,(e_i,\bm{p}_i)_{i=2}^N)=\\
&\left\{\begin{array}{cc}
\frac{1}{|\mathbb{M}(\mathbf{v},s_1,(e_i,\bm{p}_i)_{i=2}^N)|},&\mbox{if } v_i-\bm{p}_i(s_1)=\max_{j }\{(v_j-\bm{p}_j(s_1))^+\}\mbox{ and } r_i(s_1,(e_i,\bm{p}_i)_{i=1}^N)=1\\
0,&\mbox{otherwise}
\end{array}\right.,
\end{align*} 
for all $i \in \{1,\ldots,N\}$ and $(s_i,(e_i,\bm{p}_i)_{i=2}^N) \in S$, where $\mathbb{M}(\mathbf{v},s_1,(e_i,\bm{p}_i)_{i=2}^N):=\argmax_i\{(v_i-\bm{p}_i(s_1))\}$ and $\bm{p}_1(s_1)=s_1$.

\end{spacing}

\clearpage{}

\phantomsection

\end{spacing}

\clearpage{}
\end{document}